\documentclass[journal ]{new-aiaa}
\usepackage[utf8]{inputenc}
\usepackage{subcaption}
\usepackage{graphicx}
\usepackage{amsmath}
\usepackage{makecell} 
\usepackage{setspace}
\usepackage{comment}
\usepackage{multirow}

\usepackage{amsmath} 
\usepackage{amsfonts}
\usepackage{float} 

\usepackage{subcaption}

\usepackage{tikz}
\usetikzlibrary{shapes,arrows,calc,positioning}
\tikzstyle{bigblock} = [draw, fill=blue!20, rectangle, 
    minimum height=6em, minimum width=8em]
\tikzstyle{medblock} = [draw, fill=blue!20, rectangle, 
    minimum height=4em, minimum width=4em]    
\tikzstyle{mux} = [draw, fill=black!20, rectangle, 
    minimum height=5em, minimum width=0.1em]    
\tikzstyle{smallblock} = [draw, fill=blue!20, rectangle, 
    minimum height=2em, minimum width=3em]
    
\tikzstyle{data_block} = [draw, fill=green!20, rectangle, 
    minimum height=2em, minimum width=3em]
\tikzstyle{ops_block} = [draw, fill=blue!20, rectangle, 
    minimum height=2em, minimum width=3em]    
\tikzstyle{est_block} = [draw, fill=red!20, rectangle, 
    minimum height=2em, minimum width=3em]    
    
\tikzstyle{sum} = [draw, fill=blue!20, circle, node distance=1cm,minimum height=0.5cm]
\tikzstyle{signal} = [coordinate]
\tikzstyle{pinstyle} = [pin edge={to-,thin,black}]
\tikzstyle{block} = [draw, fill=blue!20, rectangle, 
    minimum height=3em, minimum width=9em]
\tikzstyle{blockS} = [draw, fill=blue!20, rectangle, 
    minimum height=3em, minimum width=4em]    
\tikzstyle{input} = [coordinate]
\tikzstyle{output} = [coordinate]
\usetikzlibrary{matrix}

\usetikzlibrary{positioning}
\usetikzlibrary{math}

\usepackage{hyperref}
\usepackage{xcolor}
\hypersetup{
    colorlinks,
    linkcolor={blue!100!black},
    citecolor={blue!50!black},
    urlcolor={blue!80!black}
}

\newcommand{\ihat}[1]{\ensuremath{\hat \imath_{\rm #1} }}
\newcommand{\jhat}[1]{\ensuremath{\hat \jmath_{\rm #1} }}
\newcommand{\khat}[1]{\ensuremath{\hat k_{\rm #1} }}

\newcommand{\bc}{\begin{center}}
\newcommand{\ec}{\end{center}}
\newcommand{\benum}{\begin{enumerate}}
\newcommand{\eenum}{\end{enumerate}}
\newcommand{\nn}{\nonumber}
\newcommand{\matl}{\left[ \begin{array}}
\newcommand{\matr}{\end{array} \right]}

\renewcommand{\matl}{\begin{bmatrix}}
\renewcommand{\matr}{\end{bmatrix}}

\newcommand{\matls}{\left[ \begin{smallmatrix}}
\newcommand{\matrs}{\end{smallmatrix} \right]}
\newcommand{\isdef}{\stackrel{\triangle}{=}}

\newcommand{\vect}[1]{\overset{\rightharpoonup}{#1}}

\newcommand{\rmA}{{\rm A}}
\newcommand{\rmB}{{\rm B}}

\newcommand{\rmE}{{\rm E}}
\newcommand{\rmF}{{\rm F}}

\newcommand{\rmL}{{\rm L}}

\newcommand{\rmP}{{\rm P}}

\newcommand{\rmT}{{\rm T}}

\newcommand{\rmc}{{\rm c}}

\newcommand{\rme}{{\rm e}}

\newcommand{\rmh}{{\rm h}}

\newcommand{\rmm}{{\rm m}}

\newcommand{\rmp}{{\rm p}}

\newcommand{\BBR}{{\mathbb R}}

\newcommand{\SF}{{\mathcal F}}

\newcommand{\SO}{{\mathcal O}}

\newcommand{\SW}{{\mathcal W}}

\newcommand{\SZ}{{\mathcal Z}}

\usepackage{enumitem,amssymb}
\newlist{todolist}{itemize}{2}
\setlist[todolist]{label=$\square$}
\usepackage{pifont}
\newcommand{\framedot}[2]{\stackrel{{\rm #1}\bullet}{#2}}

\usepackage[colorinlistoftodos,textwidth=1 in,textsize=footnotesize]{todonotes}
\title{Feedback Linearization-based Guidance Law \\ for Guaranteed Interception}
\title{Feedback Linearization-Based Guidance with \\ Zero-Dynamics Correction for Guaranteed Interception}
\author{
Alexander Dorsey%
\footnote{Graduate Research Assistant, Department of Mechanical Engineering, University of Maryland, Baltimore County, 1000 Hilltop Circle, Baltimore, MD 21250.} and
Ankit Goel\footnote{Assistant Professor, Department of Mechanical Engineering, University of Maryland, Baltimore County, 1000 Hilltop Circle, Baltimore, MD 21250.}
}

\begin{document}
\maketitle

\begin{abstract}
This paper develops a guidance law for nonlinear interception using input-output feedback linearization (IOL). 
The engagement between a pursuer and an evader is modeled using point-mass dynamics, and a baseline IOL-based guidance law is constructed by regulating the angular rates of the line-of-sight (LOS) vector. 
While this approach yields stable input-output behavior, it does not constrain the internal (zero) dynamics of the system, which can result in non-intercepting trajectories despite successful regulation of the LOS rates.
To address this limitation, a modified IOL-based guidance law is proposed that incorporates a correction mechanism to enforce convergence of the range. 
The resulting formulation ensures that LOS alignment corresponds to a closing trajectory, thereby enabling convergence of the pursuer to the evader for a broad class of initial engagement geometries. 
The proposed method retains the computational simplicity and real-time implementability of feedback linearization while improving closed-loop performance relative to classical guidance laws.
Extensive Monte Carlo simulations over a wide range of initial conditions are conducted to evaluate the proposed method. 
The results demonstrate improved reliability, reduced miss distance, and consistent convergence compared to the baseline IOL and classical proportional navigation.
\end{abstract}

\section{Introduction}

Missile guidance remains a challenging problem due to nonlinear engagement dynamics, nonminimum phase behavior arising from nose-mounted sensors and tail-fin actuation, actuator constraints, and uncertain aerodynamic loading. 
The proportional guidance law remains the most widely used approach \cite{zarchan2012tactical} due to its simplicity and low computational cost. 
However, it is derived under restrictive kinematic assumptions that are often violated in realistic engagement scenarios, limiting its performance and robustness.

Optimal control formulations have been explored to obtain guidance laws with guaranteed interception \cite{chen2010optimal}. 
While these methods provide a systematic framework, they typically require solving complex optimization problems offline and are therefore not well suited for real-time implementation. 
Polynomial and time-to-go-based guidance laws have also been proposed \cite{tahk2019augmented}, but such approaches do not explicitly address closed-loop stability, and thus cannot guarantee interception in the presence of modeling uncertainty or disturbances.

This paper develops a guidance framework based on input-output feedback linearization (IOL) to address these limitations. 
The IOL framework has been previously applied to launch vehicle guidance \cite{burchett2005feedback} and kinematic interception problems \cite{alkaher2014guidance}, and provides a systematic approach for transforming nonlinear dynamics into linear input-output relationships \cite{isidori1985nonlinear, portella2024circumventing, delgado2024adaptive}. 
In this work, the interception problem is formulated using point-mass dynamics for both the pursuer and the evader, and the guidance law is constructed using line-of-sight (LOS) measurements.

A baseline IOL-based guidance law is first derived by regulating the LOS angular rates. 
While this formulation yields stable input-output behavior, it does not uniquely determine the evolution of the internal dynamics, and can result in trajectories that diverge from the evader despite satisfying the LOS rate regulation objective. 
This limitation is rooted in the zero dynamics of the feedback-linearized system, which govern the internal evolution of the engagement when the LOS rates are regulated to zero.

The main contribution of this paper is the development of a modified IOL-based guidance law that resolves this deficiency. 
Specifically, a switching-based correction mechanism, termed the \textit{Closing Alignment Toggle Scheme (CATS)}, is introduced to enforce convergence of the range variable, thereby constraining the zero dynamics and ensuring that LOS alignment corresponds to a closing trajectory. 
The resulting guidance law enables convergence of the pursuer to the evader for a broad class of initial engagement geometries, while retaining the structure and real-time implementability of the IOL framework. 
Extensive Monte Carlo simulations are used to evaluate performance across a wide range of initial conditions and demonstrate improved reliability relative to baseline IOL and classical proportional navigation.

The paper is organized as follows. 
Section \ref{interception_dynamics} reviews the three-dimensional engagement dynamics. 
Section \ref{sec:guidance} develops the IOL-based guidance law and analyzes its properties. 
Section \ref{sec:MCS} evaluates the performance of the proposed method and compares it with classical proportional navigation. 
Finally, Section \ref{sec:conclusion} concludes the paper with a summary of results.

\section{Interception Dynamics Modeling}
\label{interception_dynamics}

This section develops a nonlinear model of the three-dimensional interception dynamics between a pursuer and an evader.
The objective is to obtain a control-affine state-space representation of the engagement suitable for guidance design. 
A detailed derivation of the underlying kinematic relationships can be found in \cite{kabamba2014fundamentals}.

The engagement geometry is formulated using an inertial reference frame and a line-of-sight (LOS) frame aligned with the relative position vector. 
The resulting kinematic relationships describe the evolution of the range and LOS angles and provide a geometrically meaningful representation of the relative motion.

Both the pursuer and the evader are modeled as point-mass systems with dynamics expressed in terms of speed, flight-path angle, and heading angle. 
These kinematic and dynamic components are combined to obtain a nonlinear, control-affine state-space model of the engagement.
This model forms the basis for the input-output feedback linearization guidance law developed in Section \ref{ssec:IOL} \cite{portella2024circumventing}.

\subsection{Engagement Geometry and Reference Frames}

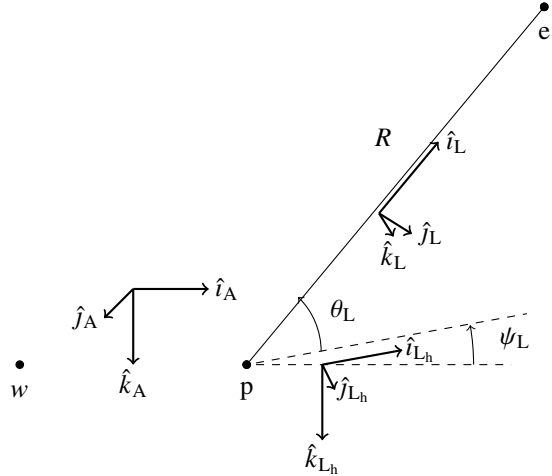
\begin{figure}[h]
    \centering    
    {
    \begin{tikzpicture}

    \tikzmath
    {
        \xF1 = -1.5;
        \yF1 = 1;
        \angF1 = 0;
    } 
    \draw [thick, ->] (\xF1, \yF1, 0) -- +({cos(\angF1)}, {sin(\angF1)}, 0) 
        node[xshift=7, yshift=0] {$\ihat{A}$};
    \draw [thick, ->] (\xF1, \yF1) -- +({cos(\angF1-90)}, {sin(\angF1-90)}) 
        node[xshift=0, yshift=-7] {$\khat{A}$};
    \draw [thick, ->] (\xF1, \yF1, 0) -- +(0,0,1) 
        node[xshift=-7, yshift=0] {$\jhat{A}$};

    \tikzmath
    {
        \xF1 = 1;
        \yF1 = 0;
        \angF1 = -30;
    } 
    \draw [thick, ->] (\xF1, \yF1, 0) -- +({cos(\angF1)}, 0, {sin(\angF1)}) 
        node[xshift=7, yshift=0] {$\ihat{L_\rmh}$};        
    \draw [thick, ->] (\xF1, \yF1, 0) -- +({-sin(\angF1)}, 0, {cos(\angF1)}) 
        node[xshift=7, yshift=0] {$\jhat{L_\rmh}$};
    \draw [thick, ->] (\xF1, \yF1,0) -- +(0,-1,0) 
        node[xshift=0, yshift=-7] {$\khat{L_\rmh}$};

    \tikzmath
    {
        \xF1 = 1.75;
        \yF1 = 2;
        \angF1 = -45;
    } 
    \draw [thick, ->] (\xF1, \yF1, 0) -- +({cos(\angF1)}, {cos(30)}, {sin(\angF1)+sin(30)}) 
        node[xshift=7, yshift=0] {$\ihat{L}$};        
    \draw [thick, ->] (\xF1, \yF1, 0) -- +({-sin(\angF1)}, 0, {cos(\angF1)}) 
        node[xshift=7, yshift=0] {$\jhat{L}$};
    \draw [thick, ->] (\xF1, \yF1,0) -- +(0,{-1+sin(30)},{-cos(60)}) 
        node[xshift=0, yshift=-7] {$\khat{L}$};

    \node at (0,0,0) (P) {};    
    \draw [fill=black] (P) circle [radius=0.050];

    \node at ({5*cos(\angF1)}, {5*cos(30)}, {5*(sin(\angF1)+sin(30))}) (E) {};  
    \draw [fill=black] (E) circle [radius=0.050];

    \draw [fill=black] (-3,0) circle [radius=0.050] node[xshift=0, yshift=-10] {$w$};
    
    \draw[->] (P.center) 
            node[xshift=00, yshift=-10] {$\rmp $} circle [radius=0.050]
            -- 
            (E.center) node[xshift=00, yshift=-10] {$\rme  $} circle [radius=0.050];

    \draw[-, dashed] (P) -- +(3.5, 0);
    \draw[-, dashed] (P) -- +({3.5*cos(-30)}, 0,{3.5*sin(-30)});

    \draw [->] (3,0) arc [radius=3, start angle=0, end angle= 10] node[xshift=17, yshift=-5] {$\psi_\rmL$};
    \draw [->] ({1*cos(10)},{1*sin(10)}) arc [radius=1, start angle=0, end angle= 46] node[xshift=17, yshift=-5] {$\theta_\rmL$};

    \node at (1.8,3) (R) {$R$};  
    \end{tikzpicture}
    }
    \caption{Interception geometry for the pursuer and the evader.}
    \label{fig:3Dinterception_geometry}
\end{figure}

Figure \ref{fig:3Dinterception_geometry} shows the three-dimensional interception geometry between the pursuer $\rmp$ and the evader $\rme$. 
The relative position vector from the pursuer to the evader specifies line of sight (LOS) direction and the range. 
The line-of-sight direction is used to define the LOS frame $\rm F_L$ that provides a convenient representation of relative kinematics and naturally separates the radial and angular components of motion.
We consider an Earth-fixed inertial frame $\rm F_A$.
For simplicity, the $\khat A$ axis of $\rm F_A$ is assumed to be aligned with the local gravity direction, while $\ihat A$ and $\jhat A$ point toward the local north and east directions, respectively, at a convenient reference location such as the launch point of the pursuer $p$.

The LOS frame $\rm F_L$ is obtained from the reference frame $\rm F_A$ through a sequence of two Euler rotations.
The first rotation is about $\khat A$ by the azimuth angle $\psi_\rmL$, resulting in the intermediate frame $\rm F_{L_h}$ such that the projection of the LOS vector on the $\ihat A-\jhat A$ plane is aligned with the $\ihat {L_h}$ vector. 
The second rotation is about $\jhat {L_h}$ by the elevation angle $\theta_\rmL$, resulting in the LOS frame $\rm F_L$ such that $\ihat L$ is along the LOS vector.
The frame vectors thus satisfy 
\begin{align}
    \SF_{\rm L_h}
        =
            \SO_3(\psi_\rmL) \SF_\rmA,        
    \quad 
    \SF_{\rm L}
        &=
            \SO_2(\theta_\rmL) \SF_{\rm L_h},
    \label{eq:frame_relations_LhA}
\end{align}
where 
\begin{align}
    \SF_\rmA 
        \isdef 
            \matl 
                \ihat{A} \\
                \jhat{A} \\
                \khat{A}
            \matr,
    \quad 
    \SF_{\rm L_h}
        \isdef 
            \matl 
                \ihat{{\rm L_h}} \\
                \jhat{{\rm L_h}} \\
                \khat{{\rm L_h}}
            \matr,
    \quad 
    \SF_\rmL 
        \isdef 
            \matl 
                \ihat{L} \\
                \jhat{L} \\
                \khat{L}
            \matr,
    \label{eq:vector_of_frame_vectors}
\end{align}
and, 
for $\phi \in \BBR,$ the orientation matrices are given by
\begin{align}
    \SO_3(\phi)
        &\isdef
            \matl
                \cos \phi & \sin \phi & 0 \\
                -\sin \phi & \cos \phi & 0 \\
                0 & 0 & 1
            \matr,
    \qquad
    \SO_2(\phi)
        \isdef
            \matl
                \cos \phi & 0 & -\sin \phi \\
                0 & 1 & 0 \\
                \sin \phi & 0 & \cos \phi
            \matr
\end{align}

The time variation of the LOS frame is characterized by its angular velocity relative to frame $\rm F_A$.
This angular velocity captures the rotation of the LOS direction due to changes in azimuth and elevation and is given by
\begin{align}
    \vect \omega_{\rm L/A}
        &=
            \dot \psi_\rmL \khat{\rm L_h}
            +
            \dot \theta_\rmL \jhat \rmL
        =
            - \dot \psi_\rmL \sin(\theta_\rmL) \ihat \rmL
            + \dot \theta_\rmL \jhat \rmL
            + \dot \psi_\rmL \cos(\theta_\rmL) \khat \rmL,
            \label{eq:omega_LA}
\end{align}
where $\khat{\rm L_h} = \sin(\theta_\rmL) \ihat \rmL + \cos(\theta_\rmL) \khat \rmL.$
Using this angular velocity, the inertial derivative of the LOS frame vector $\ihat \rmL$ can be computed using the transport theorem, that is, 
\begin{align}
    \framedot{A}{\ihat \rmL}
        &=
            \vect \omega_{\rm L/A} \times \ihat \rmL
        =
            \dot \psi_\rmL \cos(\theta_\rmL) \jhat \rmL
            - \dot \theta_\rmL \khat \rmL.
            \label{eq:ihatL_dot}
\end{align}

Similarly, the pursuer and evader body frames, $\rm F_P$ and $\rm F_E$ respectively, are defined such that the $\ihat P$ and $\ihat E$ vectors are aligned with velocity vectors of the pursuer $\rmp$ and the evader $\rme$ with respect to $\rm F_A.$
The resulting frames, therefore, satisfy
%
\begin{align}
    \SF_{\rm P_h}
        &=
            \SO_3(\psi_\rmp) \SF_\rmA, 
    \quad 
    \SF_{\rm P}
        =
            \SO_2(\theta_\rmp) \SF_{\rm P_h},
    \label{eq:frame_relations_PhA}
    \\
    \SF_{\rm E_h}
        &=
            \SO_3(\psi_\rme) \SF_\rmA, 
    \quad 
    \SF_{\rm E}
        =
            \SO_2(\theta_\rme) \SF_{\rm E_h},
    \label{eq:frame_relations_EhA}
\end{align}
where $\SF_{\rm P_h}, \SF_{\rm P}, \SF_{\rm E_h},$ and $\SF_{\rm E}$ denote the vectors of frame unit vectors defined similarly to \eqref{eq:vector_of_frame_vectors}.
Note that $\psi_\rmp$ and $\psi_\rme$ are the \textit{heading angles} of the pursuer and the evader, respectively, while $\theta_\rmp$ and $\theta_\rme$ are their corresponding \textit{flight-path angles}.
Furthermore, the velocities of the pursuer and evader relative to a fixed point in $\rm F_A$ satisfy 
$\vect v_{\rmp/w/\rmA} = V_\rmp \ihat P$ and
$\vect v_{\rme/w/\rmA} = V_\rme \ihat E,$ respectively, 
where $V_\rmp$ and $V_\rme$ are the magnitudes of the velocity of the pursuer and the evader, respectively. 

\subsection{Relative Motion Kinematics}

The relative position vector from the pursuer to the evader is aligned with the LOS and is given by
$\vect r_{\rm e/p} 
        = 
            R \ihat \rmL,$
where $R$ denotes the range.
The velocity of the evader relative to the pursuer with respect to $\rm F_A$ is then given by 
\begin{align}
    \vect v_{\rm e/p/\rmA}
        &=
            \framedot{A}{\vect r}_{\rm e/p}
        =
            \dot R \ihat \rmL
            +
            R \framedot{A}{\ihat \rmL}
        =
            \dot R \ihat \rmL
            +
            R
            \left(
                \dot \psi_\rmL \cos(\theta_\rmL) \jhat \rmL
                - \dot \theta_\rmL \khat \rmL
            \right)
        =
            \matl 
                \dot R &                
                R\dot \psi_\rmL \cos(\theta_\rmL) &
                - R\dot \theta_\rmL                 
            \matr
            \SF_\rmL
    \label{eq:v_epa_1}
\end{align}

The same relative velocity can alternatively be expressed as the difference between the evader and pursuer inertial velocities, that is, 
Consequently, the velocity of the evader relative to the pursuer with respect to $\rm F_A$  satisfies
\begin{align}
    \vect v_{\rm e/p/\rmA}
        &=
            \vect v_{\rme/w/\rmA}
            -
            \vect v_{\rmp/w/\rmA}
        =
            V_\rme \ihat \rmE
            -
            V_\rmp \ihat \rmp.
            \label{eq:vepA_diff}
\end{align}
Using \eqref{eq:frame_relations_LhA}, \eqref{eq:frame_relations_PhA}, and \eqref{eq:frame_relations_EhA}, it follows that 
\begin{align}
    \ihat{\rmp} 
        &= 
            e_1^\rmT \SF_\rmp
        =
            e_1^\rmT \SO_2(\theta_\rmp) \SO_3(\psi_\rmp) \SO_3(\psi_\rmL)^\rmT \SO_2(\theta_\rmL)^\rmT \SF_{\rm L},
            \label{eq:F_L_2_i_P}
    \\
    \ihat{\rmE} 
        &= 
            e_1^\rmT \SF_\rme
        =
            e_1^\rmT \SO_2(\theta_\rme) \SO_3(\psi_\rme) \SO_3(\psi_\rmL)^\rmT \SO_2(\theta_\rmL)^\rmT \SF_{\rm L},
            \label{eq:F_L_2_i_E}
\end{align}
and thus 
\begin{align}
    \vect v_{\rm e/p/\rmA}
        &=
            \left[
            V_\rme 
            e_1^\rmT \SO_2(\theta_\rme) \SO_3(\psi_\rme) \SO_3(\psi_\rmL)^\rmT \SO_2(\theta_\rmL)^\rmT 
            -
            V_\rmp 
            e_1^\rmT \SO_2(\theta_\rmp) \SO_3(\psi_\rmp) \SO_3(\psi_\rmL)^\rmT \SO_2(\theta_\rmL)^\rmT 
            \right]
            \SF_{\rm L}.
    \label{eq:v_epa_2}
\end{align}

Using \eqref{eq:v_epa_1} and \eqref{eq:v_epa_2}, it follows that 
\begin{align}
    \matl 
        \dot R \\
        R\dot \psi_\rmL \cos(\theta_\rmL) \\
        - R\dot \theta_\rmL                 
    \matr
        =
            \SO_2(\theta_\rmL)
            \SO_3(\psi_\rmL)
            \left[
            V_\rme             
            \SO_3(\psi_\rme)^\rmT  
            \SO_2(\theta_\rme)^\rmT 
            -
            V_\rmp             
            \SO_3(\psi_\rmp)^\rmT
            \SO_2(\theta_\rmp)^\rmT
            \right]
            e_1,
    \label{eq:LOS_dynamics}
\end{align}
which can be written in the scalar form as 
\begin{align}
    \dot R
        &=
            V_\rme 
            \Big(
                \sin(\theta_\rme)\sin(\theta_\rmL)
                +
                \cos(\theta_\rmL)\sigma_\rme
            \Big)
            -
            V_\rmp
            \Big(
                \sin(\theta_\rmp)\sin(\theta_\rmL)
                +
                \cos(\theta_\rmL)\sigma_\rmp
            \Big),
    \label{eq:Rdot_3D}
    \\
    \dot \psi_\rmL
        &=
            \dfrac{
                V_\rme \cos(\theta_\rme)\sin(\psi_\rme-\psi_\rmL)
                +
                V_\rmp \cos(\theta_\rmp)\sin(\psi_\rmL-\psi_\rmp)
            }{
                R \cos(\theta_\rmL)
            },
    \label{eq:psidot_3D_0}
    \\
    \dot \theta_\rmL
        &=
            -\dfrac{
                V_\rmp 
                \Big(
                    \cos(\theta_\rmL)\sin(\theta_\rmp)
                    -
                    \sin(\theta_\rmL)\sigma_\rmp
                \Big)
                -
                V_\rme
                \Big(
                    \cos(\theta_\rmL)\sin(\theta_\rme)
                    -
                    \sin(\theta_\rmL)\sigma_\rme
                \Big)
            }{
                R
            },
        \label{eq:psidot_3D}
\end{align}
where
    $\sigma_{\rmp}
        \isdef
            \cos(\theta_\rmp)\cos(\psi_\rmL-\psi_\rmp),$
    $\sigma_{\rmE}
        \isdef
            \cos(\theta_\rme)\cos(\psi_\rme-\psi_\rmL).$
Equations \eqref{eq:Rdot_3D}-\eqref{eq:psidot_3D} describe the evolution of the engagement in terms of radial and angular motion.
The range dynamics capture the closing or diverging behavior along the LOS, while the angular dynamics describe the rotation of the LOS vector due to relative motion.
These equations reveal that the LOS angular rates depend on both the relative geometry and the velocity alignment between the pursuer and evader.
Note that when $\psi_\rmL = \psi_\rme = \psi_\rmp = 0,$ the engagement reduces to a planar interception problem, presented in \cite{dorsey2025feedbacklinearizationbasedguidancelaw}, thereby providing consistency with existing planar dynamics. 

These equations describe the evolution of the relative range and LOS angles as functions of the pursuer and evader headings and speeds.
In the subsequent section, these nonlinear dynamics are exploited to construct an input-output linearizing guidance law that directly regulates the LOS rates to enforce interception.

\subsection{Pursuer and Evader Dynamics}
In this work, both the pursuer and the evader are modeled as point-mass bodies. 
The motion of each vehicle is described in terms of its speed $V$, flight-path angle $\theta$, and heading angle $\psi$.

Consider a point-mass body $\rmB$. Let $\rmF_\rmA$ be an inertial frame with $\khat \rmA$ aligned with the gravity vector, and let $w$ denote a point with zero inertial acceleration. The body-fixed frame $\rmF_\rmB$ is defined such that $\ihat \rmB$ is aligned with the velocity vector $\vect v_{\rmc/w/\rmA}$.
The body velocity magnitude is denoted by $V_\rmB$, and the orientation of $\rmF_\rmB$ relative to $\rmF_\rmA$ is parameterized by the heading angle $\psi$ and the flight-path angle $\theta$.

The translational dynamics of the point-mass body are given by
\begin{align}
    \dot V_\rmB
        &=
            - g \sin \theta
            +
            \frac{1}{m_\rmB}
            \big(
                T_{\rmB,x}
                -
                D_\rmB
            \big),
    \label{eq:Vdot}
    \\
    \dot \psi
        &=
            \frac{
                T_{\rmB,y}
            }{
                m_\rmB V_\rmB \cos \theta
            },
    \label{eq:psidot}
    \\
    \dot \theta
        &=
            -\frac{1}{V_\rmB}
            \left(
                g \cos \theta
                +
                \frac{T_{\rmB,z}}{m_\rmB}
            \right),
    \label{eq:thetadot}
\end{align}
where, $m_\rmB$ is the body mass, $g$ is the gravitational acceleration, $D_\rmB$ is the drag force acting along $-\ihat B$, and $T_{\rmB,x}, T_{\rmB,y}, T_{\rmB,z}$ denote the components of the control force expressed in the body-fixed frame.

These dynamics are used to model both the pursuer and the evader. 
In particular, the pursuer states $(V_p, \theta_p, \psi_p)$ and the evader states $(V_e, \theta_e, \psi_e)$ evolve according to the above equations with their respective parameters and inputs.
In this work, the drag acting on both the pursuer and the evader is modeled as described in \cite{islam2022minimum}.

\subsection{Engagement Dynamics in State-Space Form}

To facilitate the application of input-output feedback linearization, the interception dynamics derived in the previous section are recast into a state-space representation. 
This form explicitly separates the natural dynamics of the engagement from the control influence of the pursuer and the evader dynamics, and provides a convenient framework for characterizing relative degree, zero dynamics, and control effectiveness.

The state vector is chosen to reflect the relative engagement geometry and pursuer kinematics, while the evader motion is treated as an exogenous disturbance input. 
Specifically, define the state, disturbance, and control vectors as
\begin{align}
    x 
        &\isdef
        \begin{bmatrix}
            R &
            \theta_\rmL &
            \psi_\rmL &
            V_\rmp &
            \theta_\rmp &
            \psi_\rmp
        \end{bmatrix}^\rmT,
    \\
    w
        &\isdef
        \begin{bmatrix}
            V_\rme &
            \theta_\rme &
            \psi_\rme
        \end{bmatrix}^\rmT,
    \\
    u
        &\isdef
        \begin{bmatrix}
            n_{y,\rmp} &
            n_{z,\rmp}
        \end{bmatrix}^\rmT .
\end{align}

The state vector $x$ consists of the range $R$, the LOS elevation and azimuth angles $(\theta_\rmL,\psi_\rmL)$, and the pursuer speed. heading, and flight path angle $(V_\rmp,\theta_\rmp,\psi_\rmp)$. 
The disturbance vector $w$ contains the evader speed and orientation variables, which influence the relative dynamics but are not controlled by the pursuer. 
The control input $u$ consists of the local lateral and vertical acceleration commands available to the pursuer.
Using \eqref{eq:LOS_dynamics} and \eqref{eq:Vdot}-\eqref{eq:thetadot}, the engagement dynamics can be written in the affine nonlinear form
\begin{align}
    \dot x
        =
            f(x,w)
            +
            g(x) u,
    \label{eq:SS_form}
\end{align}
where $f(x,w)$ represents the engagement dynamics and $g(x)$ describes how the pursuer control inputs influence the state.
Specifically,
\begin{align}
    f(x,w)
        \isdef
        \begin{bmatrix}
            \dot R \\[0.4em]
            \dot \theta_\rmL \\[0.4em]
            \dot \psi_\rmL \\[0.4em]
            \dot V_\rmp \\[0.4em]
            \dot \theta_\rmp \\[0.4em]
            \dot \psi_\rmp
        \end{bmatrix}
        =
        \begin{bmatrix}
            V_\rme
            \Big(
                \sin(\theta_\rme)\sin(\theta_\rmL)
                +
                \cos(\theta_\rmL)\sigma_\rme
            \Big)
            -
            V_\rmp
            \Big(
                \sin(\theta_\rmp)\sin(\theta_\rmL)
                +
                \cos(\theta_\rmL)\sigma_\rmp
            \Big)
            \\[0.8em]
            -\dfrac{
                V_\rmp
                \Big(
                    \cos(\theta_\rmL)\sin(\theta_\rmp)
                    -
                    \sin(\theta_\rmL)\sigma_\rmp
                \Big)
                -
                V_\rme
                \Big(
                    \cos(\theta_\rmL)\sin(\theta_\rme)
                    -
                    \sin(\theta_\rmL)\sigma_\rme
                \Big)
            }{
                R
            }
            \\[1.0em]
            \dfrac{
                V_\rme \cos(\theta_\rme)\sin(\psi_\rme-\psi_\rmL)
                +
                V_\rmp \cos(\theta_\rmp)\sin(\psi_\rmL-\psi_\rmp)
            }{
                R \cos(\theta_\rmL)
            }
            \\[1.0em]
            - g \sin(\theta_\rmp)
            +
            \dfrac{
                T_{\rmp,x}
                -
                D_\rmp
            }{
                m_\rmp
            }
            \\[1.0em]
            -\dfrac{g \cos(\theta_\rmp)}{V_\rmp}
            \\[1.0em]
            0
        \end{bmatrix},
\end{align}
and
\begin{align}
    g(x)
        \isdef
        \begin{bmatrix}
            0 & 0 \\
            0 & 0 \\
            0 & 0 \\
            0 & 0 \\
            0 & -\dfrac{1}{V_\rmp} \\
            -\dfrac{1}{V_\rmp \cos(\theta_\rmp)} & 0
        \end{bmatrix}.
\end{align}

The evader dynamics are modeled as an exogenous nonlinear system
\begin{align}
    \dot w
        =
            \SW(w),
\end{align}
where the evader dynamics 
\begin{align}
    \SW(w)
        \isdef
        \begin{bmatrix}
            \dfrac{T_\rme - D_\rme}{m_\rme}
            -
            g \sin(\theta_\rme)
            \\[0.8em]
            -\dfrac{
                g \cos(\theta_\rme)
                +
                n_{z,\rmE}
            }{
                V_\rme
            }
            \\[1.0em]
            \dfrac{
                n_{y,\rmE}
            }{
                V_\rme \cos(\theta_\rme)
            }
        \end{bmatrix}.
\end{align}
Note that the evader state $w$ acts as an exogenous input to the engagement dynamics, where $\SW(w)$ captures the evader translational and orientation dynamics, including thrust, drag, and maneuver accelerations. 
The function $\SW$ allows modeling of aggressive and adversarial evader maneuvers.

The objective of the engagement problem is to design a guidance law that ensures interception of the evader. 
Specifically, the guidance law must drive the range $R$ to zero with a sufficient closing velocity so that the pursuer crosses the interception point. 
To accomplish this objective, we use input-output feedback linearization to transform the engagement dynamics into a linear system with respect to the range variable.

\section{Guidance Design and Analysis}
\label{sec:guidance}
This section develops a guidance law for the interception problem using the input-output feedback linearization (IOL) framework. 
The objective is to regulate the angular rates of the line-of-sight (LOS) vector to zero, thereby enforcing alignment between the pursuer velocity and the LOS direction, which is a necessary condition for interception.

The engagement dynamics derived in Section II are first transformed using input-output linearization to obtain a control law that directly regulates the LOS angular rates. 
The resulting closed-loop system is then analyzed to characterize its validity conditions and internal (zero) dynamics. 
In particular, it is shown that regulation of the LOS rates does not uniquely determine the evolution of the range, thereby allowing non-intercepting trajectories.

To address this limitation, a modification to the baseline IOL guidance law is introduced to ensure convergence to the intercepting solution. 
The properties of the resulting guidance law are analyzed and compared with classical proportional guidance.

\subsection{Input--Output Linearization of LOS Dynamics}
\label{ssec:IOL} 
In this work, we linearize the engagement dynamics from the control input to the angular rates of the line-of-sight (LOS) vector. 
Specifically, the dynamics from $u$ to $\dot{\psi}_{\rmL}$ and $\dot{\theta}_{\rmL}$ are linearized. 
This choice is motivated by the geometric interpretation provided by the classical interception triangle in guidance theory. 
In both cases, the LOS vector relative to the pursuer approaches a constant direction, which implies that the LOS angular rates satisfy 
$\dot{\psi}_{\rmL} \to 0$ and $\dot{\theta}_{\rmL} \to 0$. 
By linearizing the dynamics from $u$ to these LOS angular rates, the interception problem can be recast as a regulation problem in which controllers are designed to drive $\dot{\psi}_{\rmL}$ and $\dot{\theta}_{\rmL}$ to zero using standard linear control techniques.

Define
\begin{align}
    y
        \isdef
            h(x)
        \isdef
            \begin{bmatrix}
                \dot \psi_\rmL \\
                \dot \theta_\rmL
            \end{bmatrix}.
\end{align}
This choice of output corresponds to directly shaping the LOS kinematics rather than the LOS angles themselves.
Note that the relative degree of the system with respect to each output, that is, $\dot \psi_\rmL$ and $ \dot \theta_\rmL,$ is one, indicating that the control inputs appear explicitly in the first derivative of each output. 

Following the general framework in \cite{delgado2023circumventingunstablezerodynamics}, the input-output linearized representation of the LOS rate dynamics can be written in the linear time-invariant form
\begin{align}
    \dot \xi
        &=
            A_\rmc \xi
            +
            B_\rmc 
            (\alpha(x) + \beta(x) u),
    \\
    y
        &=
            C_\rmc \xi,
\end{align}
where $\xi$ is the linearizable state, and the matrices $A_\rmc$, $B_\rmc$, and $C_\rmc$ are given by
\begin{align}
    A_\rmc
        &\isdef
            0_{2\times 2}
    \quad
    B_\rmc
        \isdef
            I_2,
    \quad
    C_\rmc
        \isdef
            I_2,
\end{align}
and the functions $\alpha(x,w)$ and $\beta(x)$ are given by
\begin{align}
    \alpha(x,w)
        &\isdef
            L_f h(x)
        =
            \dfrac{\partial h(x)}{\partial x} f(x,w)
            \in \mathbb{R}^{2 \times 1}, 
    \\
    \beta(x)
        &\isdef
            L_g h(x)
         =
            \dfrac{\partial h(x)}{\partial x} g(x)
            \in \mathbb{R}^{2 \times 2}.
\end{align}

Assuming that $\beta(x)$ is invertible, the nonlinear feedback law
\begin{align}
    u(x,w)
        =
            \beta(x)^{-1}
            \left(
                -\alpha(x,w)
                +
                v
            \right),
    \label{eqn:IOL_control_dotPsi}
\end{align}
yields the closed-loop dynamics
\begin{align}
    \dot \xi
        &=
            A_\rmc \xi
            +
            B_\rmc v.
\end{align}
Consequently, under exact feedback linearization, the LOS rates can be regulated arbitrarily via the internal control signal $v$.
%
For example, if the internal control signal $v$ is chosen as
\begin{align}
    v
        =
            \begin{bmatrix}
                -k_{\dot \psi} \dot \psi_\rmL \\
                -k_{\dot \theta} \dot \theta_\rmL
            \end{bmatrix},
            \label{eq:GuidanceControlLaw_v}
\end{align}
where $k_{\dot \psi} , k_{\dot \theta} > 0$,
then each LOS rate channel satisfies exponentially stable first-order dynamics, 

\subsection{Validity of the Linearizing Control Law}

Note that, defining $\Delta\psi \isdef \psi_\rmp-\psi_\rmL, $
\begin{align}
    \beta(x)
        =
        \frac{1}{R}
        \begin{bmatrix}
            -\cos(\Delta\psi)
            &
            -\sin(\theta_\rmp)\sin(\Delta\psi)
            \\
            -\sin(\theta_\rmL)\sin(\Delta\psi)
            &
            \cos(\theta_\rmp)\cos(\theta_\rmL)
            + \sin(\theta_\rmp)\sin(\theta_\rmL)\cos(\Delta\psi)
        \end{bmatrix}.
\end{align}
and
\begin{align}
    \beta^{-1}(x)
        \frac{1}{R \det\beta(x)}
        \begin{bmatrix}
            \cos(\theta_\rmp)\cos(\theta_\rmL)
            +\sin(\theta_\rmp)\sin(\theta_\rmL)\cos(\Delta\psi)
            &
            \sin(\theta_\rmp)\sin(\Delta\psi)
            \\
            \sin(\theta_\rmL)\sin(\Delta\psi)
            &
            -\cos(\Delta\psi)
        \end{bmatrix},
\end{align}
and
\begin{align}
    \det\beta(x)
        =
        -\frac{
            \sin(\theta_\rmp)\sin(\theta_\rmL)
            +
            \cos(\theta_\rmp)\cos(\theta_\rmL)\cos(\Delta\psi)
        }{R^2}.
\end{align}

Noting that
\begin{align}
    \ihat{\rmP}^\rmT \ihat{\rmL}
        =
        \sin(\theta_\rmp)\sin(\theta_\rmL)
        +
        \cos(\theta_\rmp)\cos(\theta_\rmL)\cos(\Delta\psi),
\end{align}
it follows that
\begin{align}
    \det \beta(x)
        =
        -\frac{\ihat{\rmP}^\rmT \ihat{\rmL}}{R^2}.
\end{align}
Therefore, the inverse is
\begin{align}
    \beta^{-1}(x)
        =
        \frac{R}{\ihat{\rmP}^\rmT \ihat{\rmL}}
        \begin{bmatrix}
            \cos(\theta_\rmp)\cos(\theta_\rmL)
            +
            \sin(\theta_\rmp)\sin(\theta_\rmL)\cos(\Delta\psi)
            &
            \sin(\theta_\rmp)\sin(\Delta\psi)
            \\
            \sin(\theta_\rmL)\sin(\Delta\psi)
            &
            -\cos(\Delta\psi)
        \end{bmatrix}.
\end{align}
Thus, the input-output linearizing control law is well defined as long as the pursuer's velocity vector is not orthogonal to the LOS vector, that is,  $\ihat{\rmP}^\rmT \ihat{\rmL} \neq 0.$ 

\subsection{Zero Dynamics and Non-Intercepting Trajectories}

Since $\xi = y \in \BBR^2,$ the remaining state variables constitute the zero-dynamics state. 
Defining
\begin{align}
    \eta 
        \isdef 
        \begin{bmatrix}
            R &
            V_\rmp &
            \theta_\rmp &
            \psi_\rmp
        \end{bmatrix}^\rmT
        \in \BBR^4,
\end{align}
the dynamics of $\eta$ are given by
\begin{align}
    \dot{\eta} = \SZ(\eta,\xi),
\end{align}
where the dynamics $\SZ$ is called the zero dynamics or the internal dynamics of the system. 
The linearized state $\xi$ is regulated using the input-output linearization (IOL) control law. 
However, the evolution of $\eta$ is governed by the equation above. 
Since the control law is designed to directly regulate $\xi$, the state $\eta$ evolves according to the internal dynamics of the system.

\begin{figure}[h!]
    \centering
    \includegraphics[width=0.5\linewidth]{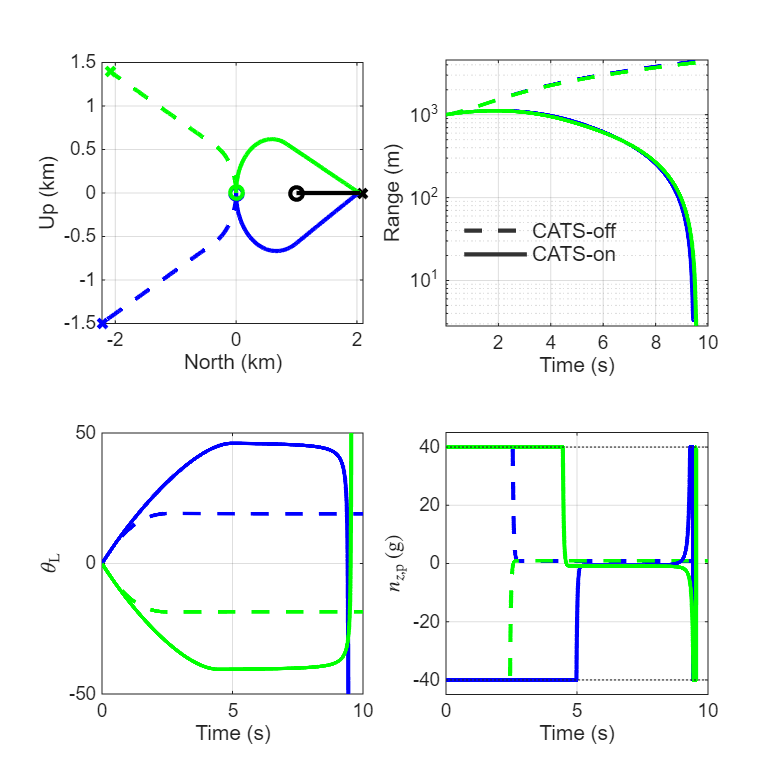}
    \caption{Bistable LOS dynamics under baseline and CATS-modified guidance.}
    \label{fig:PrePostLogicFigure}
\end{figure}

We emphasize that the proposed guidance law guarantees $h(x) \to 0$, but does not explicitly regulate the line-of-sight angles $\psi_\rmL$ and $\theta_\rmL$.
As a result, in certain engagement geometries, the pursuer trajectory may satisfy $\dot \psi_\rmL \to 0$ and $\dot \theta_\rmL \to 0$ while the relative distance $R$ diverges.
This corresponds to a degenerate case in which the pursuer velocity becomes aligned with the LOS direction but points away from the evader, leading to non-intercepting trajectories.
This behavior is illustrated in Figure \ref{fig:PrePostLogicFigure} as the dashed lines.

The regulation of the LOS angular rates does not uniquely determine the evolution of the range variable $R$. 
Consequently, even if the linearized state $\xi$ converges to zero, the remaining state $\eta$ may evolve along different zero-dynamics trajectories.

From the range kinematics,
\begin{align}
    \dot R
        =
        \ihat{\rmL}^{\rmT}
        \vect v_{\rme/w/\rmA}
        -
        V_\rmp
        \ihat{\rmP}^{\rmT}
        \ihat{\rmL},
    \label{eq:R_dot_zero_dynamics}
\end{align}
where $\ihat{\rmL}$ is the unit line-of-sight (LOS) vector and $\ihat{\rmP}$ is the unit vector along the pursuer velocity direction.
Equation \eqref{eq:R_dot_zero_dynamics} shows that the sign of the inner product $\ihat{\rmP}^{\rmT}\ihat{\rmL}$ determines whether the pursuer has a closing or diverging velocity component along the LOS.
When the LOS rates are regulated to zero, that is, 
\begin{align}
    \dot{\psi}_\rmL \to 0,
    \qquad
    \dot{\theta}_\rmL \to 0,
\end{align}
the LOS direction approaches a constant vector. 
However, two distinct motion patterns are possible.

\begin{enumerate}
\item 
\textbf{Closing motion:}
If $\ihat{\rmP}^{\rmT}\ihat{\rmL}>0,$ the pursuer velocity has a component directed toward the evader.
In this case, the term $-V_\rmp \ihat{\rmP}^{\rmT}\ihat{\rmL}$ contributes negatively to $\dot R$, and the range decreases provided the pursuer has sufficient closing-speed authority.

\item 
\textbf{Diverging motion:}
If $\ihat{\rmP}^{\rmT}\ihat{\rmL}<0,$ the pursuer velocity is directed away from the evader along the LOS.
In this case, the pursuer contributes positively to $\dot R$, and the range increases even though the LOS rates are driven to zero.
\end{enumerate}

Therefore, the manifold defined by
    $\dot{\psi}_\rmL = 0,$
and 
    $\dot{\theta}_\rmL = 0$
contains two branches of the zero dynamics, one corresponding to closing trajectories and another corresponding to non-intercepting trajectories.

The modification introduced in \eqref{eq:IOL_control_dotPsi_mod} removes the non-intercepting branch by ensuring that the pursuer velocity remains aligned toward the evader, that is, 
\begin{align}
    \ihat{\rmP}^{\rmT}\ihat{\rmL} > 0.
\end{align}
As a result, the pursuer always maintains a closing velocity component along the LOS.
Consequently, once the LOS angular rates converge to zero, the remaining internal dynamics drive the range variable $R$ monotonically toward zero, guaranteeing interception.

\subsection{Closing Alignment Toggle Scheme (CATS)}

To mitigate the issue of velocity LOS alignment, a simple geometric correction logic is incorporated to ensure that the pursuer's velocity remains directed toward the evader. This method is called Closing Alignment Toggle Scheme (CATS).
Using the velocity $\vect v_{\rmp/w/\rmA}$ of the pursuer with respect to the inertial frame $\rm F_A$
and the relative position vector $\vect r_{\rm p/e}$ as
the modified control law is then given by
\begin{align}
    u(x,w)
        =
        \begin{cases}
            \beta(x)^{-1} \big( -\alpha(x,w) + v \big),
                & 
            \ihat{\rmP}^\rmT \ihat{\rmL} < 0
            \\[0.6em]
            -\beta(x)^{-1} \big( -\alpha(x,w) + v \big),
                & 
            \ihat{\rmP}^\rmT \ihat{\rmL}
            \ge 0,
        \end{cases}
    \label{eq:IOL_control_dotPsi_mod}
\end{align}
which ensures convergence of the relative distance while preserving the line-of-sight rate regulation.
This correction mechanism effectively enforces a directional consistency condition, preventing the controller from stabilizing non-intercepting equilibrium configurations associated with unfavorable velocity alignment. The effects of the modification are seen in \ref{fig:PrePostLogicFigure} as the Solid Lines, which shows the correction of the bistable effect by forcing the velocity to align appropriately with the relative position vector.

\subsection{Comparison with Proportional Guidance}
\label{sec:comparison_pg}

The proposed input-output linearization (IOL) guidance law differs fundamentally from classical proportional guidance (PG), also known as proportional navigation, in both its control structure and its dependence on measured quantities.

Proportional guidance commands the pursuer's turn rate proportional to the line-of-sight (LOS) angular rate. 
In its classical form, proportional guidance commands the pursuer's lateral acceleration proportional to the line-of-sight (LOS) angular rates. 
Using the LOS angles $(\psi_L, \theta_L)$, the commanded accelerations can be expressed as
\begin{align}
    n_{y,p} &= N V_p \dot{\psi}_L, \\
    n_{z,p} &=  N V_p \dot{\theta}_L,
\end{align}
where $N$ is the navigation constant.
Thus, PG uses LOS rate as a measured signal to indirectly influence the engagement geometry.

In contrast, the proposed IOL-based guidance law treats the LOS angular rates $(\dot{\psi}_L, \dot{\theta}_L)$ as system outputs and directly regulates them via feedback linearization.
This yields closed-loop dynamics in which the LOS rates evolve according to user-defined stable dynamics. 
As a result, the guidance objective is explicitly formulated as
$\dot{\psi}_L \to 0, $ and $ \dot{\theta}_L \to 0,$
rather than being achieved indirectly through proportional feedback.

A key distinction between the two approaches lies in how they handle the internal (zero) dynamics.
In PG, convergence of the LOS rate to zero is not explicitly enforced through system inversion, and interception relies on favorable engagement geometry and sufficient closing velocity. 
Consequently, PG may exhibit non-intercepting trajectories under certain initial conditions.
On the other hand, in the IOL framework, the LOS rate dynamics are explicitly linearized and regulated. 
However, this exposes the system's internal zero dynamics, which can admit both closed and diverging trajectories. 
The proposed closing alignment toggle scheme (CATS) ensures that the system evolves along the closing branch of the zero dynamics, thereby guaranteeing interception.

The two approaches also differ in their sensing requirements. 
Proportional guidance requires only measurements of the LOS rate and, in some implementations, relative range or closing velocity. 
These quantities are typically obtained from onboard seekers or tracking filters, making PG attractive for practical implementation.
On the contrary, the IOL-based guidance law requires additional state information to evaluate the nonlinear functions $\alpha(x,w)$ and $\beta(x)$. 
In particular, the implementation requires measurements or estimates of the pursuer states $(V_p, \theta_p, \psi_p)$, the LOS angles $(\theta_L, \psi_L)$ and their rates, as well as information about the evader motion $(V_e, \theta_e, \psi_e)$.

While these requirements are more demanding than those of PG, many of these quantities are either directly measured (e.g., LOS angles and rates) or available through onboard navigation and state estimation systems. 
In modern guidance systems, such information is routinely obtained through inertial measurement units, GPS, and sensor fusion algorithms.
Therefore, the increased measurement requirements do not pose a fundamental limitation, but rather reflect the more detailed model-based structure of the IOL approach.

In summary, proportional guidance provides a simple and robust heuristic based on LOS rate feedback, whereas the proposed IOL-based guidance offers a systematic control-theoretic framework that directly regulates LOS dynamics and explicitly accounts for the underlying system structure.
The trade-off between the two approaches lies in simplicity versus model-based precision and control authority.

\section{Numerical Simulations} \label{sec:MCS}

This section evaluates the performance of the proposed input-output linearization (IOL) guidance law through numerical simulations of three-dimensional interception scenarios. 
The objective is to assess the ability of the proposed approach to achieve reliable interception across a range of engagement geometries, including both rear-aspect and front-aspect configurations, under evasive and non-evasive target maneuvers.

The performance of the IOL-based guidance law is compared with that of classical proportional guidance (PG) under identical initial conditions and acceleration constraints.
Particular attention is given to the effect of the zero dynamics on interception performance and the role of the closing alignment toggle scheme (CATS) in ensuring convergence to the intercepting solution. 
Monte Carlo simulations are used to quantify robustness in terms of miss distance and failure rate.

\subsection{Simulation Setup}

The engagement dynamics are modeled using three-dimensional point-mass representations for both the pursuer and the evader. 
The evader is modeled with a constant thrust of $50~\rm kN$, a mass of $10{,}000~\rm kg$, and gravity-induced normal acceleration. 
The pursuer is powered by a constant thrust of $15~\rm kN$ and is controlled through lateral and vertical acceleration commands.

Two guidance laws are evaluated: the proposed line-of-sight (LOS)-rate-based IOL guidance law, and the classical proportional guidance (PG) law. 
The PG law is implemented with a fixed navigation gain of $3$, while the IOL gains are selected as $k_{\dot{\psi}} = 50$ and $k_{\dot{\theta}} = 50$ for all simulations.

Each Monte Carlo experiment consists of $10{,}000$ trials with initial conditions uniformly sampled from the ranges listed in Table \ref{tab:MC_IC_summary}. 
An engagement is declared successful if the miss distance at intercept is less than $10~\rmm$ \cite{aerospace10020155}. 
For each trial, the interception time, miss distance, and closing velocity are recorded.

Four engagement configurations are considered: rear-aspect and front-aspect encounters with non-maneuvering evaders, and corresponding cases with maneuvering evaders. 
In the maneuvering scenarios, the evader executes a discrete $10g$ acceleration maneuver at a randomized time, with maneuver direction selected randomly in both course and elevation.

To isolate intrinsic guidance performance, sensor noise and measurement uncertainty are not included. 
In all cases, the pursuer is initialized with a higher speed than the evader to represent terminal-phase interception conditions. 
The Monte Carlo sampling procedure follows \cite{PalumbHoming}.

\begin{figure}[h!]
    \centering
    \includegraphics[width=0.7\linewidth]{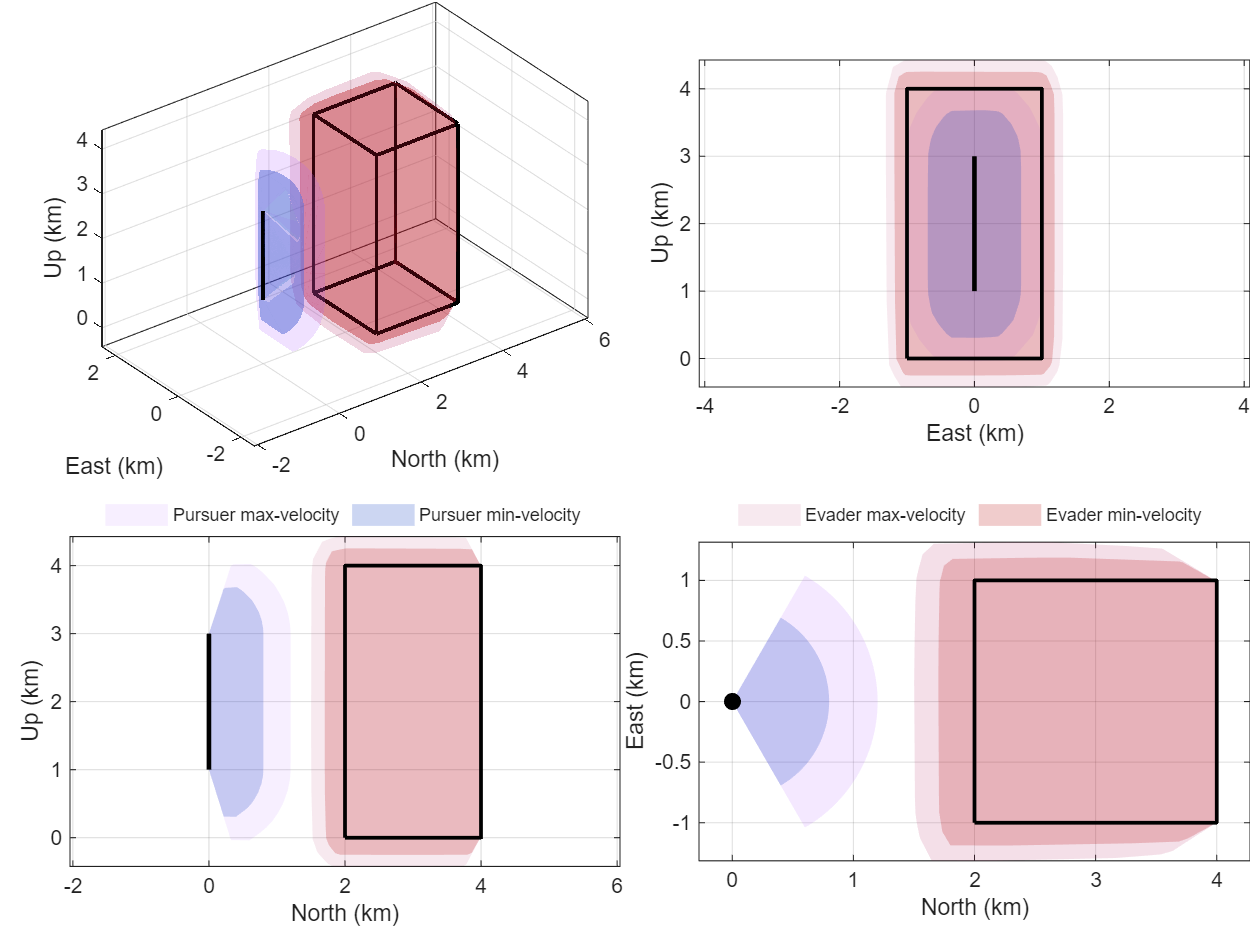}
    \caption{Initial conditions and reachable regions over two seconds.}
    \label{fig:IC_FA_Cartoon}
\end{figure}

\begin{table*}[h]
\centering
\small
\renewcommand{\arraystretch}{1.15}
\setlength{\tabcolsep}{6pt}
\resizebox{\textwidth}{!}{%
\begin{tabular}{|c|c|c|c|c|c|c|c|}
\hline
\textbf{Case} & \textbf{Vehicle} &
Velocity ($\rm m/s$) & $\theta$ (deg) & $\psi$ (deg) &
North (km) & East (km) & Altitude (km) \\
\hline

RA & Pursuer & (400,600) & (-60,60) & (-60,60) & 0 & 0 & (1,3) \\
RA & Evader  & (150,250) & (-60,60) & (-60,60) & (2,4) & (2,4) & (0,4) \\
\hline

FA & Pursuer & (400,600) & (-60,60) & (-60,60) & 0 & 0 & (1,3) \\
FA & Evader  & (150,250) & (-60,60) & (140,220) & (2,4) & (-1,1) & (0,4) \\
\hline

FAE & Pursuer & (800,1100) & (-20,20) & (-20,20) & 0 & 0 & 10 \\
FAE & Evader  & (300,600)  & (-15,15) & (175,195) & 10 & (-1,1) & 10 \\
\hline

RAE & Pursuer & (800,1100) & (-20,20) & (-20,20) & -10 & 0 & 10 \\
RAE & Evader  & (300,600)  & (-15,15) & (-20,20) & 0 & (-1,1) & 10 \\
\hline

\end{tabular}}
\caption{Ranges of uniformly distributed initial conditions for Monte Carlo simulations. The cases where the evader is only subject to air resistance and gravity are the Rear Aspect (RA) and Front Aspect (FA). Maneuvering cases (FAE, RAE) include evasion times of $(0,7)$ s and $(0,25)$ s, respectively.}
\label{tab:MC_IC_summary}
\end{table*}

\subsection{Nominal Engagements}

The nominal engagement cases provide a baseline comparison of the guidance laws in the absence of evader maneuvers. 
These scenarios isolate the intrinsic behavior of each guidance strategy and highlight differences in convergence rate, accuracy, and variability.

Table \ref{tab:MC_nominal} summarizes the Monte Carlo results for rear-aspect (RA) and front-aspect (FA) geometries. 
In both cases, the line-of-sight (LOS)-rate-based IOL guidance law achieves shorter interception times, significantly smaller mean miss distances, and reduced variability compared to proportional guidance. 
In the rear-aspect case, the IOL law reduces the average miss distance by more than an order of magnitude and lowers the failure rate from $2.2982\%$ to $0.3997\%$. 
In the front-aspect case, where high closing velocity reduces the available time-to-go, the IOL law maintains lower miss-distance dispersion and achieves a lower failure rate ($0.0500\%$ vs $0.2498\%$), indicating improved robustness under more demanding conditions.

\begin{table*}[h]
\centering
\resizebox{\textwidth}{!}
{
\begin{tabular}{|c|c|ccc|ccc|}
\hline
\textbf{Case} & \textbf{Metric}
& \multicolumn{3}{c|}{\textbf{Line-of-sight-based IOL}}
& \multicolumn{3}{c|}{\textbf{Proportional Guidance}} \\
\hline
 &  & Time (s) & Miss (m) & Closing Vel. (m/s)
    & Time (s) & Miss (m) & Closing Vel. (m/s) \\
\hline
\multirow{3}{*}{RA}
& Avg & 11.375 & 1.7034 & -458.77 & 12.487 & 17.326 & -482.48 \\
& Std & 2.6485 & 30.705 & 78.972 & 3.0894 & 151.33 & 93.521 \\
& Fail & \multicolumn{3}{c|}{0.3997\%} 
       & \multicolumn{3}{c|}{2.2982\%} \\
\hline
\multirow{3}{*}{FA}
& Avg & 5.5511 & 0.34702 & -642.6 & 5.9844 & 0.46434 & -600.22 \\
& Std & 1.3251 & 1.169 & 93.143 & 1.6286 & 8.4125 & 114.9 \\
& Fail & \multicolumn{3}{c|}{0.0500\%} 
       & \multicolumn{3}{c|}{0.2498\%} \\
\hline
\end{tabular}}
\caption{Monte Carlo results for nominal rear-aspect (RA) and front-aspect (FA) engagement scenarios.}
\label{tab:MC_nominal}
\end{table*}

Representative engagement trajectories are shown in Figures ~\ref{fig:4x4_Example_MCS_All}, while the empirical cumulative distribution functions in Figure ~\ref{fig:ecdf_2_by_3} further illustrate the improved consistency of the IOL guidance law across the Monte Carlo trials.

These results are consistent with the zero-dynamics analysis presented in Section~II. 
Regulation of the LOS angular rates alone does not uniquely determine the evolution of the relative distance, since both closing and diverging trajectories satisfy $\dot{\psi}_L \to 0$ and $\dot{\theta}_L \to 0$. 
The improved performance of the IOL guidance law indicates that it more effectively aligns the pursuer velocity with the LOS direction, thereby biasing the system toward the closing branch of the zero dynamics and accelerating convergence of the relative distance.

\subsection{Maneuvering-Evader Engagements}

We next evaluate the performance of the guidance laws in the presence of evasive target maneuvers. 
These scenarios introduce time-varying disturbances and represent more realistic and challenging interception conditions.

Table ~\ref{tab:MC_maneuvering} summarizes the Monte Carlo results for front-aspect (FAE) and rear-aspect (RAE) engagements with maneuvering evaders. 
In the front-aspect case, where high closing velocity limits the available time-to-go, the line-of-sight (LOS)-rate-based IOL guidance law achieves zero failures across all trials while maintaining consistently low miss distances. 
In contrast, proportional guidance exhibits a failure rate of $5.3957\%$ and significantly larger miss-distance dispersion, with maximum miss distances exceeding tens of meters. 
This result highlights the IOL guidance law's ability to maintain reliable interception performance under aggressive, time-varying disturbances.

\begin{table*}[h]
\centering
\resizebox{\textwidth}{!}{
\begin{tabular}{|c|c|ccc|ccc|}
\hline
\textbf{Case} & \textbf{Metric}
& \multicolumn{3}{c|}{\textbf{Line-of-sight-based IOL}}
& \multicolumn{3}{c|}{\textbf{Proportional Guidance}} \\
\hline
 &  & Time (s) & Miss (m) & Closing Vel. (m/s)
    & Time (s) & Miss (m) & Closing Vel. (m/s) \\
\hline
\multirow{3}{*}{FAE}
& Avg & 7.5834 & 0.31731 & -1266.1 & 7.7057 & 1.9103 & -1175 \\
& Std & 1.0603 & 0.19656 & 259.75 & 1.1424 & 5.8119 & 315.96 \\
& Fail & \multicolumn{3}{c|}{0.00\%} 
       & \multicolumn{3}{c|}{5.3957\%} \\
\hline
\multirow{3}{*}{RAE}
& Avg & 17.771 & 1.1074 & -708.24 & 17.942 & 1.3947 & -681.3 \\
& Std & 3.7464 & 19.807 & 187.56 & 3.7673 & 23.47 & 160.83 \\
& Fail & \multicolumn{3}{c|}{0.2998\%} 
       & \multicolumn{3}{c|}{0.5096\%} \\
\hline
\end{tabular}}
\caption{Monte Carlo results for maneuvering-evader engagement scenarios: front-aspect (FAE) and rear-aspect (RAE).}
\label{tab:MC_maneuvering}
\end{table*}

In the rear-aspect maneuvering scenario, both guidance laws experience increased variability due to the longer engagement duration and delayed evasive maneuvers. 
Nevertheless, the IOL guidance law achieves a lower failure rate ($0.2998\%$ vs $0.5096\%$) and reduced mean miss distance, indicating improved consistency across trials. 
Although both methods exhibit occasional large-miss outliers, the IOL approach demonstrates greater robustness to late-time maneuvering.

\begin{figure}[h!]
    \centering
    \includegraphics[width=1\linewidth]{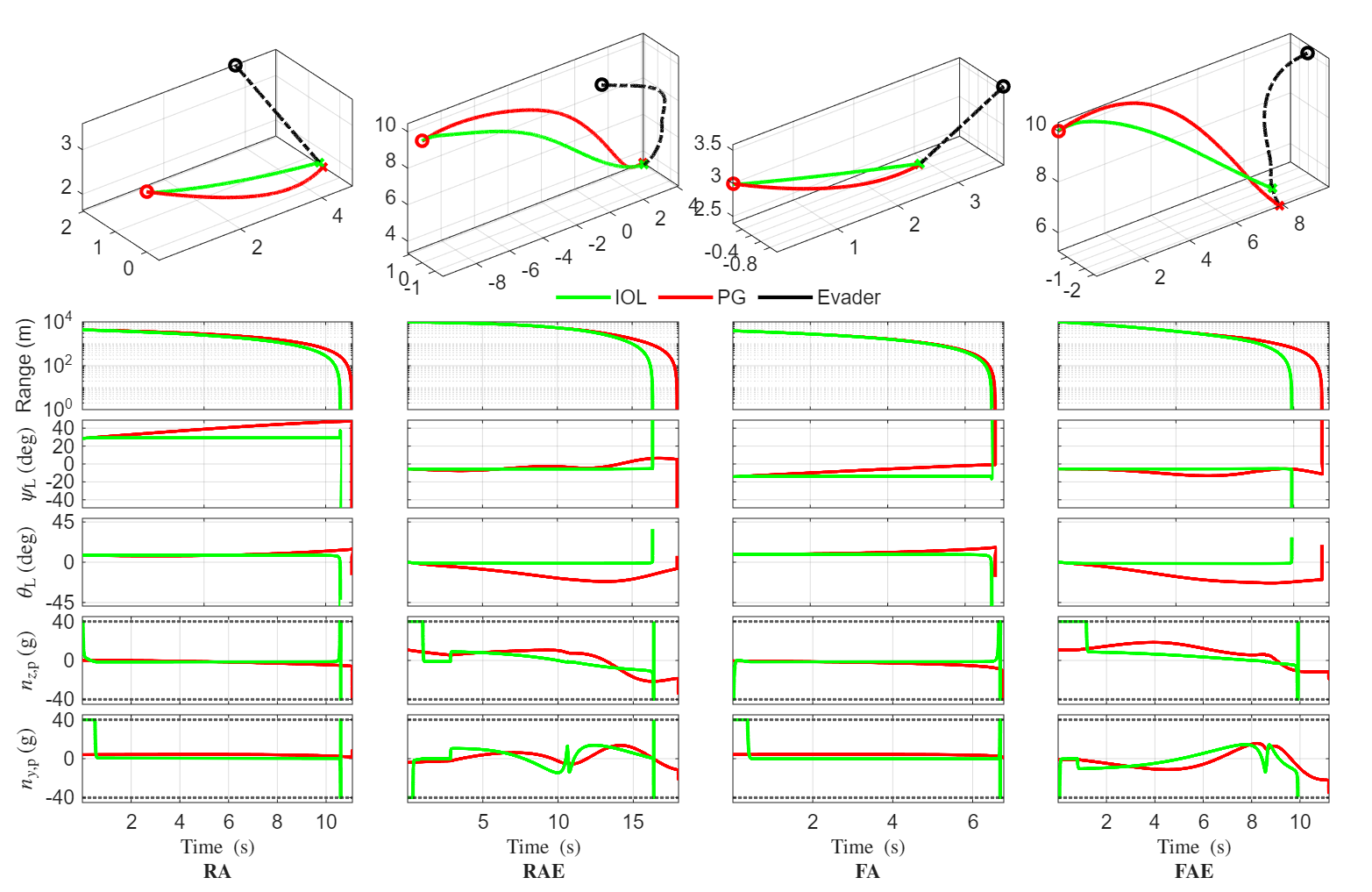}
    \caption{Representative Monte Carlo cases: trajectories, range, LOS angles, accelerations for the four engagement scenarios.}
    \label{fig:4x4_Example_MCS_All}
\end{figure}

\begin{figure}[h!]
    \centering
    \includegraphics[width=1\columnwidth]{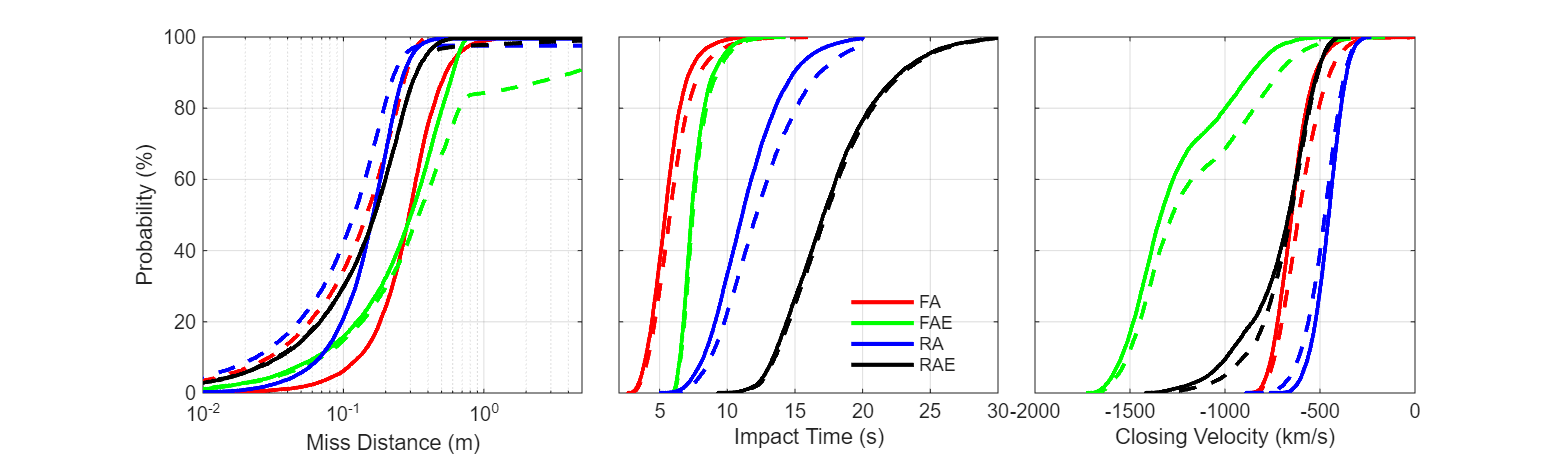}
    \caption{Empirical cumulative distribution functions (miss distance, impact time, closing velocity) for the various engagement scenarios with IOL (solid) and proportional (dashed) guidance, respectively.}
    \label{fig:ecdf_2_by_3}
\end{figure}

Representative trajectories are shown in Figure ~\ref{fig:4x4_Example_MCS_All}, while the empirical cumulative distribution functions in Figure ~\ref{fig:ecdf_2_by_3} further illustrate the reduced variability and improved robustness of the IOL guidance law.

These results are consistent with the zero-dynamics interpretation developed in Section~II. 
Evasive maneuvers introduce perturbations that can drive the system toward trajectories with unfavorable velocity-LOS alignment. 
By directly regulating the LOS angular rates while maintaining alignment between the pursuer velocity and the LOS direction, the IOL guidance law suppresses transitions toward the divergent branch of the zero dynamics. 
As a result, the pursuer preserves a closing component of velocity along the LOS even under time-varying disturbances, leading to improved interception reliability.

\section{Conclusion}
\label{sec:conclusion}
This paper developed an input-output feedback linearization (IOL) framework for guidance in three-dimensional point-mass interception scenarios.
By formulating the engagement dynamics in the line-of-sight (LOS) frame, the guidance problem was recast as the regulation of LOS angular rates.
The resulting IOL-based guidance law enables direct control over the LOS rate dynamics, providing a systematic alternative to classical proportional navigation.

Analysis of the zero dynamics revealed the existence of non-intercepting trajectories corresponding to diverging motion along the LOS, despite regulation of the LOS rates.
To address this limitation, a geometric correction mechanism, termed the closing alignment toggle scheme (CATS), was introduced to enforce a positive closing component along the LOS.
This modification eliminates the non-intercepting branch of the zero dynamics and ensures convergence to interception.

The proposed guidance law was evaluated through Monte Carlo simulations across rear-aspect and front-aspect engagement scenarios, including both evasive and non-evasive target maneuvers.
The results demonstrate that the IOL-based guidance achieves reliable interception with consistently lower miss distances compared to proportional navigation. 
In particular, the proposed approach eliminates failures in several scenarios where proportional navigation exhibits nonzero miss rates.

Overall, this work provides a principled framework for guidance design based on feedback linearization, highlights the role of zero dynamics in interception performance, and introduces a simple yet effective mechanism to guarantee convergence.
These results suggest that feedback-linearization-based approaches offer a promising direction for robust and geometrically consistent guidance in nonlinear interception problems.

\bibliography{Bib/Refs,Bib/chanl}

\end{document}